\documentclass[a4paper]{VisionStyle}
\usepackage{epsfig}
\usepackage{amssymb}
\newcommand{\Hii}{H\,{\sc ii} }
\newcommand{\Log}{\mbox{Log}}

\begin{document}

\title{ The 2--10 keV luminosity as a Star Formation Rate indicator }

\author{P.\,Ranalli\inst{1} \and A.\,Comastri\inst{2} \and 
  G.\,Setti\inst{1} } 

\institute{
  Dipartimento di Astronomia, Universit\`a di Bologna,
  via Ranzani 1, I--40127 Bologna, Italy
\and 
  INAF -- Osservatorio Astronomico di Bologna,
  via Ranzani 1, I--40127 Bologna, Italy
}

\maketitle 

\begin{abstract}
Radio and far infrared luminosities of star forming galaxies follow a
tight linear relation. Making use of BeppoSAX and ASCA observations of a
well-defined sample of star forming galaxies, we argue that a tight
linear relation holds between the 2--10 keV X-ray luminosity and both
the radio and far infrared ones. It is suggested that the hard X-ray
emission is directly related to the Star Formation Rate.
 Preliminary results obtained from deep {\em Chandra} and radio observations of
the Hubble Deep Field North show that a similar relation might hold
also at high ($0.2\lesssim z\lesssim 1.2$) redshift.

\end{abstract}

\section{Introduction}

Radio continuum and far infrared (FIR) luminosities of star forming galaxies are known
to show a {\em tight linear} relationship spanning four orders of
magnitude in luminosity
(\cite{pranalli-E3:vdK73}; \cite{pranalli-E3:dejo85}; \cite{pranalli-E3:cond92}).
This 
is interpreted as due to the presence of massive, young stars embedded in dust:
a fraction of their UV radiation is absorbed by dust grains
and  reradiated in the infrared band, while supernova explosions may
accelerate the electrons producing the observed synchrotron
emission (\cite{pranalli-E3:hp75}; \cite{pranalli-E3:helou}).
Since massive ($M\gtrsim 5 ~\rm M\sun$) stars are
short-lived, these luminosities are assumed to be indicators of the global 
Star Formation Rate (SFR) in a galaxy.
Following \cite*{pranalli-E3:cond92} and \cite*{pranalli-E3:kenn98},
the relation between SFR and radio/FIR luminosities can be written as:
\begin{eqnarray}
{\rm SFR}=\frac{L_{\rm 1.4 GHz}}{4.0\cdot 10^{28}}~ \mbox{M$\sun$/yr}
	 \label{pranalli-E3:eqsfr1}  \\
{\rm SFR}=\frac{L_{\rm FIR}}{2.2\cdot 10^{43}}~  \mbox{M$\sun$/yr}
	\label{pranalli-E3:eqsfr2}
\end{eqnarray}
with the FIR flux defined after \cite*{pranalli-E3:helou} as:
\begin{equation}
{\rm FIR}=1.26\cdot 10^{-11} (2.58 S_{60\mu} + S_{100\mu})~ \mbox{erg s$^{-1}$ cm$^{-2}$}
\end{equation}
where $L_{\rm 1.4 GHz}$ is in erg\,s$^{-1}$\,Hz$^{-1}$, $L_{\rm FIR}$ in erg$^{-1}$ and 
infrared fluxes in Jy.

Star forming galaxies are known to show luminous soft X-ray emission,
originated in hot plasmas associated to star forming regions and
galactic winds.
A non linear ($L_{\rm X} \propto L_{\rm FIR}^{0.6}$) 
and much scattered (dispersion of about 2 dex) 
relation was found between FIR and soft (0.5--3.0 keV) X-ray
luminosities measured by the {\em Einstein}
satellite (\cite{pranalli-E3:gp90}). 

BeppoSAX and ASCA observations of a sizable sample of star forming
galaxies make possible to extend the study of these relations into the
hard (2--10 keV) X-ray band.

\begin{figure*}[t]    
  \begin{center} 
      \includegraphics[width=0.47\textwidth]{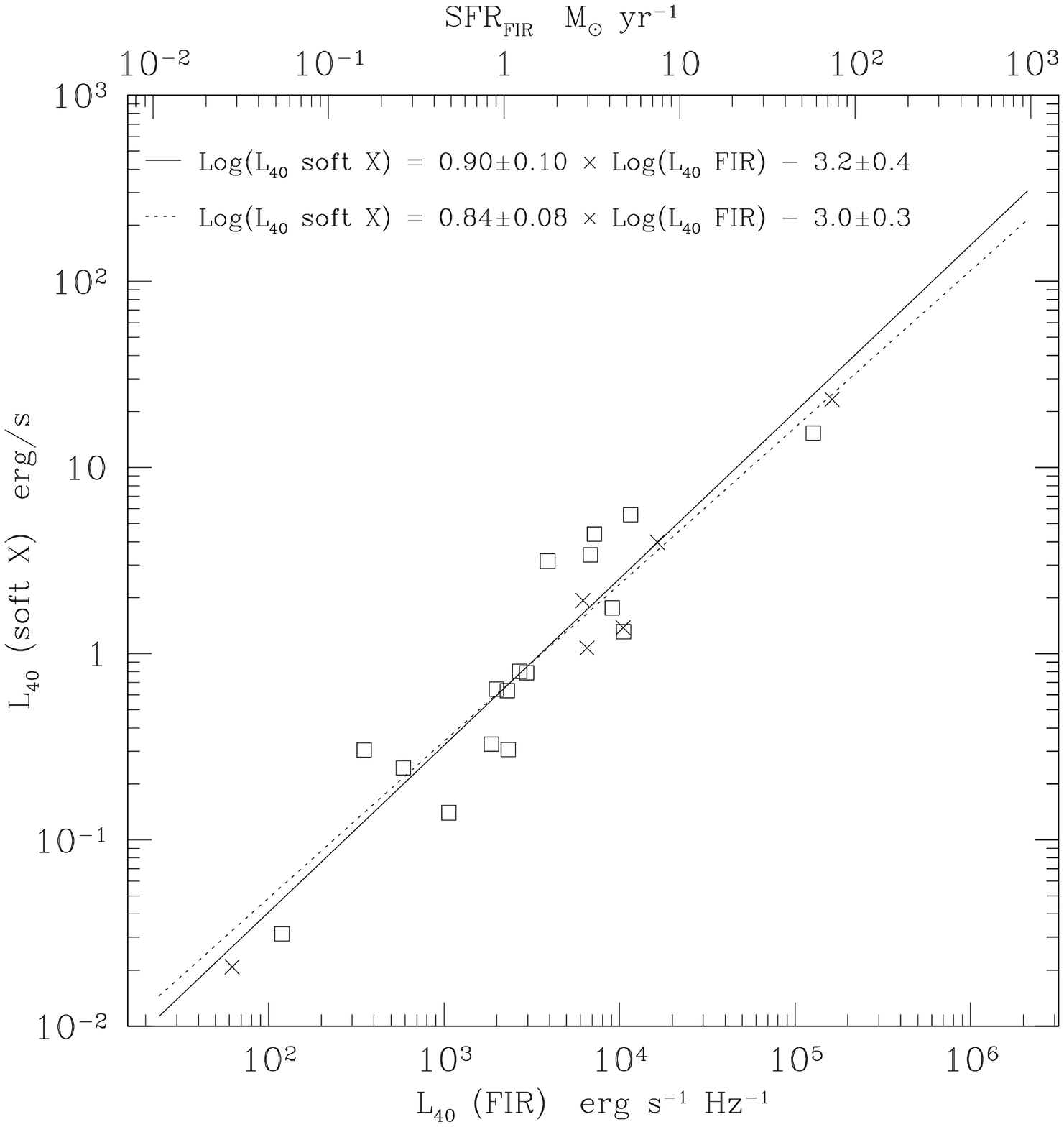}
      \hskip.8cm
      \includegraphics[width=0.47\textwidth]{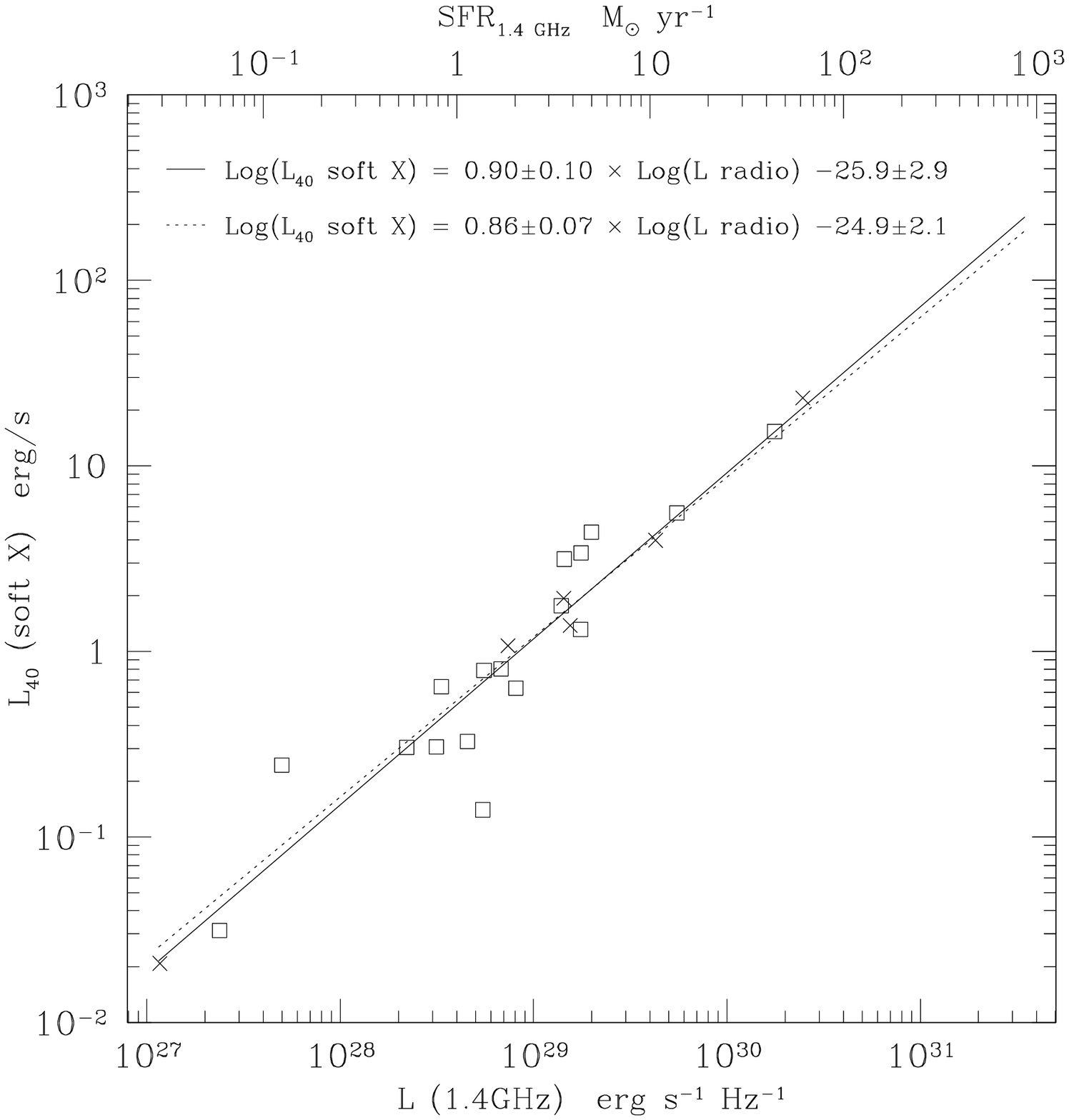}
  \end{center}
  \caption {\small The 0.5--2.0 keV luminosity of local star forming is related
  to radio and FIR ones. Straight line: best-fit relation for the local
  sample. Dotted line: best-fit for the local+supplementary samples.
  Squares: local sample. Crosses: supplementary
  sample.}
  \label{pranalli-E3_fig:fig1}
\end{figure*}

\section{The local sample}

The atlas of optical nuclear spectra by \cite*{pranalli-E3:hfs97}
(hereafter HFS97)
represents a complete spectroscopic survey of
galaxies in the Revised Shapley-Ames Catalog of Bright Galaxies by
\cite*{pranalli-E3:RSA} (RSA) with declination $\delta > 0\degr$ and
magnitude $B_T<12.5$.
Optical spectra are classified in HFS97 on the basis of line intensity ratios
according to \cite*{pranalli-E3:vo87}; galaxies with nuclear line ratios
typical of star forming systems are labeled as ``\Hii nuclei''.

We cross-correlated the HFS97 sample with the IRAS Faint
Source Catalogue (\cite{pranalli-E3:FSC}, FSC; 
notice that the FSC only covers the sky with galactic latitude 
$|b|>10\degr$ and
is complete down to limiting fluxes of 0.2\,Jy at $60\mu$ and 1.0\,Jy
at $100\mu$)
obtaining a complete homogeneous sample of 193 nearby ($z<0.01$)
star forming galaxies (hereafter the ``parent sample'')
which was then cross-correlated with
the BeppoSAX and ASCA archives. Eighteen
galaxies were detected in the 2--10 keV band with the MECS or GIS
instruments.
Four additional objects in the field of view of ASCA observations
were not detected. The 2--10 keV flux upper limits are too
loose to add any significant 
information, and thus we did not include them in the sample.
Radio (1.4\,GHz) fluxes were obtained from the Condon et
al.~(1990, 1996) catalogues.


Most of the data have already been published; in a few cases
where published data were not available, we
reduced the ASCA archival data.
Images and spectra were extracted from
the pipeline-screened event files. The images were checked against optical
(Digital Sky Survey) and, where available, radio (20\,cm) images in order to look for
possible source confusions. Fluxes were calculated in the 0.5--2.0 and
2--10 keV bands from best-fit spectra and corrected for Galactic absorption
only.
One object (\object{M33}) was not included in the sample since its
broad-band (0.5--10 keV) X-ray 
nuclear spectrum is dominated by a strong variable source (\object{M33 X-8})
identified as a black hole candidate (\cite{pranalli-E3:parmar01}).
Although M33 is identified by HFS97 as an \Hii nucleus, it has a
very low SFR ($\sim 0.009$ M$\sun$/yr) so that the spectral signatures 
related to star formation can be easily hidden by a single powerful
source.

Our sample (``local sample'') consists of the 17 galaxies listed in
Tab.~\ref{pranalli-E3_tab:smpl}. It contains 11 galaxies with
${\rm SFR}>1~ {\rm M}\sun/$yr 
out of 77   in the parent sample ($14\%$), and
5 out of 27  with ${\rm SFR}>3~ {\rm M}\sun/$yr ($19\%$).
Since the parent sample is complete within the statistical errors,
these numbers represent an estimate of our sample completeness.
All statistical tests presented here are performed on this sample unless
otherwise stated.

In the figures we also plot data 
for 6 other well-known starburst galaxies
which were not in the HFS97 survey  because they are in
the southern emisphere. On the basis of their line intensity ratios
they should be classified as \Hii nuclei.
In Tab.~\ref{pranalli-E3_tab:smpl} we label them
as ``supplementary sample''.

\section{X-rays and the Star Formation Rate}

\begin{figure*}[t]    
  \begin{center} 
      \includegraphics[width=0.47\textwidth]{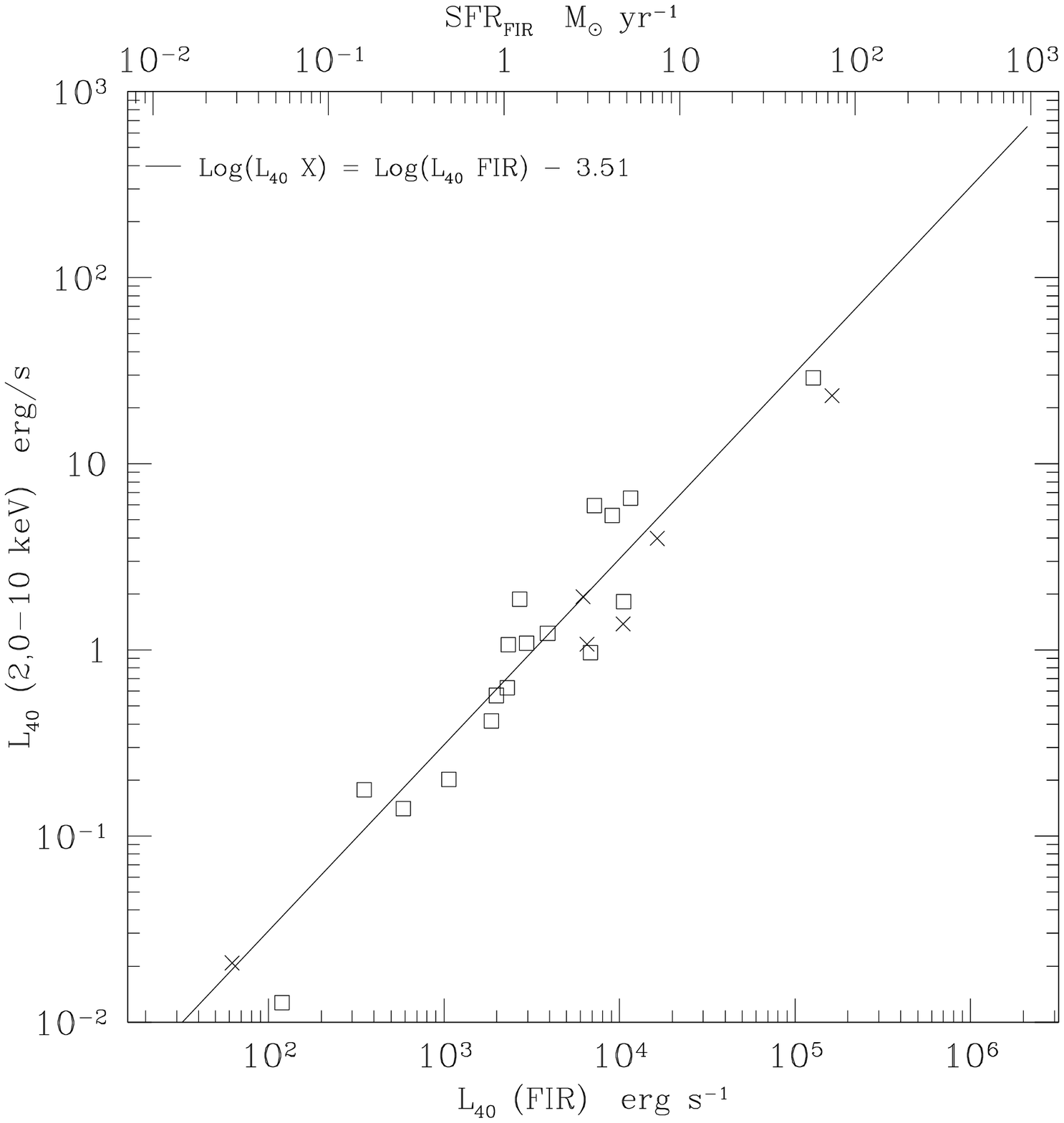}
      \hskip.8cm
      \includegraphics[width=0.47\textwidth]{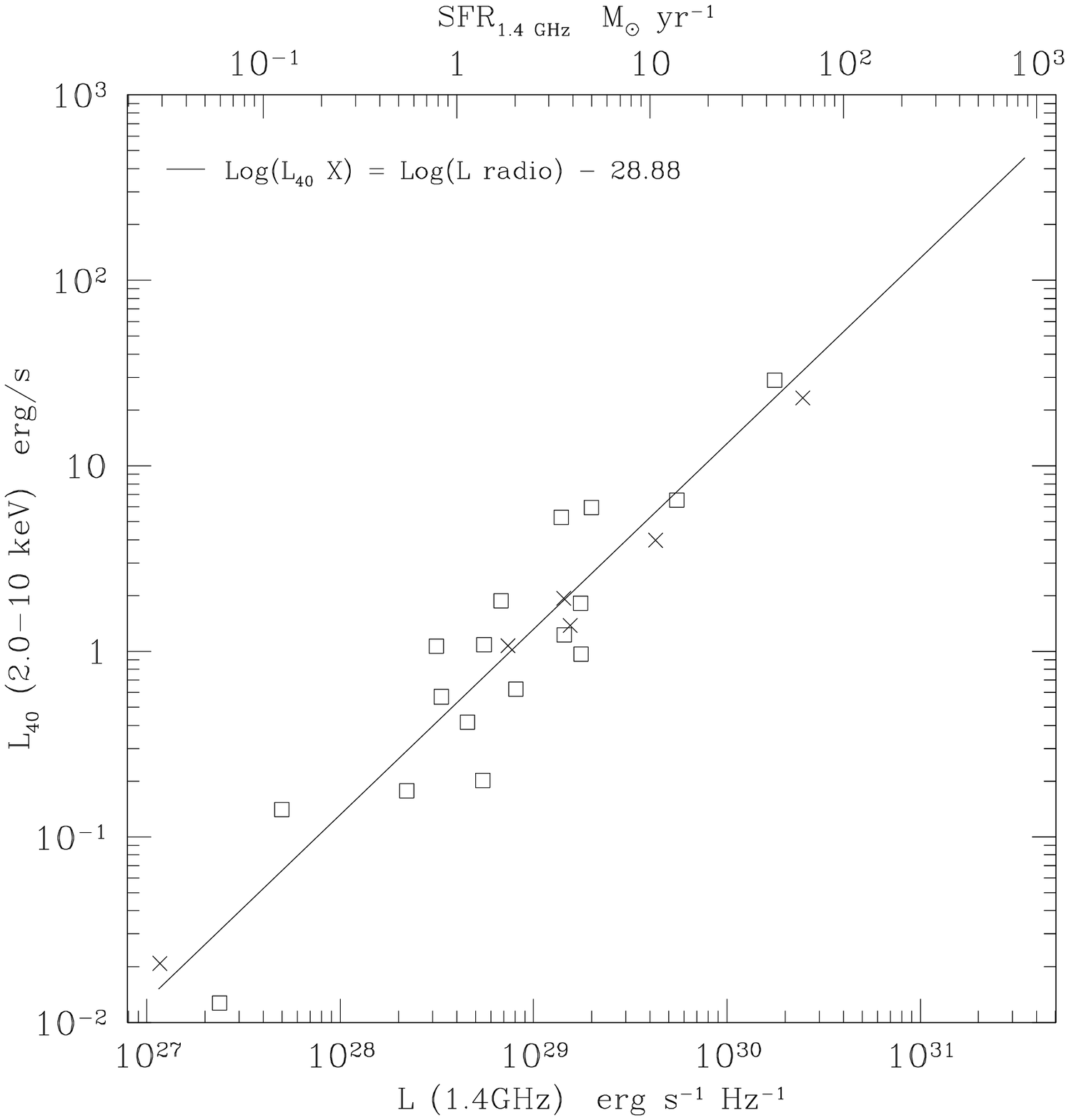}
  \end{center}
  \caption {\small The 2--10 keV luminosity of local star forming galaxies is
  linearly related to radio and FIR ones. Symbols as in fig.\,(%
  \ref{pranalli-E3_fig:fig1}). }
  \label{pranalli-E3_fig:fig2}
\end{figure*}

As a preliminary test, we perform a least-squares analysis for the
well-known radio/FIR correlation, which yields
\begin{equation}\label{pranalli-E3:radiofir}
\Log (L_{\rm FIR}) = (0.94\pm 0.09) ~\Log (L_{\rm 1.4}) 
     +16.4\pm 2.5
\end{equation}

The dispersion around the best-fit relation
is estimated as:
\begin{equation}
\delta = 1/N \cdot \sum \left| (L_{\rm obs}-L_{\rm pred})/L_{\rm pred}
\right|
\end{equation}
where $L_{\rm pred}$ is the luminosity expected from the best fit
relation and $L_{\rm obs}$ the observed one.
For the radio/FIR correlation
(eq.~\ref{pranalli-E3:radiofir})  one has $\delta \simeq 43\%$.

\begin{table}[!b]	
\caption{\small Galaxies\,in\,the\,local\,sample. All\,galaxies\,were\,observed
 by\,ASCA,\,except\,those\,marked\,with\,*\,observed\,by\,BeppoSAX.}
\label{pranalli-E3_tab:smpl}
\centering\begin{tabular}{ccc}
\multicolumn{3}{c}{\sc Local Sample}\\ 
\hline\hline

\object{M82}*     &\object{NGC\,2276} &\object{NGC\,4449}\\
\object{M101}    &\object{NGC\,2403} &\object{NGC\,4631}\\
\object{M108}    &\object{NGC\,2903} &\object{NGC\,4654}\\ 
\object{NGG\,891}  &\object{NGC\,3310}&\object{NGC\,6946}\\
\object{NGC\,1569} &\object{NGC\,3367} &\object{IC\,342}  \\
\object{NGC\,2146} &\object{NGC\,3690} \\
\\
\multicolumn{3}{c}{\sc Supplementary Sample }\\
\hline\hline
\object{NGC\,55}	&\object{NGC\,1672} &\object{NGC\,3256}\\
\object{NGC\,253}*	&\object{NGC\,1808} &\object{Antennae}\\
\end{tabular}
\end{table}

Following \cite*{pranalli-E3:helou}
we also calculate the mean ratio $q$ between the logarithms of FIR and
radio fluxes, obtaining
$q\simeq 2.05$
with a variance $\sigma\simeq 0.23$. These values can be compared with
the mean $q=2.34\pm 0.01$ for the 2809 galaxies in the IRAS 2\,Jy sample by
\cite*{pranalli-E3:yrc}. Although our sample shows a lower $q$,
this result may not be statistically significant.

As a further test we checked the soft X-ray/FIR relation 
(Fig.~\ref{pranalli-E3_fig:fig1}), finding \\
\begin{eqnarray}
\Log (L_{.5-2}) &= &(0.90\pm 0.10) ~\Log (L_{\rm FIR}) 
	-3.2\pm 0.4 \label{pranalli-E3:eqfirsx} \\
\Log (L_{.5-2}) &= &(0.90\pm 0.10) ~\Log (L_{\rm 1.4})
	 +14.1\pm 2.9 \label{pranalli-E3:eqradiosx} \medskip
\end{eqnarray}\\
with mean $\delta$'s of  58\% and  46\%, respectively.


Our result is consistent
with the $L_{0.5-4.5 \rm keV}\propto L_{\rm FIR}^{0.95\pm 0.06}$
relation found by \cite*{pranalli-E3:djf92} for normal and
starburst galaxies from the IRAS Bright Galaxy Sample, but
is inconsistent with the much flatter relationship obtained by
\cite*{pranalli-E3:gp90} for a sample of IRAS selected galaxies
($L_{0.5-3 \rm keV} \propto L_{60\mu}^{0.62\pm 0.14}$) and for
a sample of starburst/interacting galaxies
($L_{0.5-3 \rm keV} \propto L_{60\mu}^{0.70\pm 0.12}$).

The inclusion of the supplementary objects (Tab.~%
\ref{pranalli-E3_tab:smpl}) in our 
analysis would lead to a slight flattening of the slope, i.e. 
$L_{0.5-2.0 \rm keV}\propto L_{\rm FIR}^{0.84\pm 0.08}$;
likewise, if we use the $60\mu$ luminosity instead of FIR, we 
obtain $L_{0.5-2.0 \rm keV}\propto L_{60\mu}^{0.83\pm 0.10}$
with $\delta=64\%$. However these results are still consistent
with slopes derived in eq. (\ref{pranalli-E3:eqfirsx}) and
(\ref{pranalli-E3:eqradiosx}) within 1$\sigma$.

\begin{figure}[!t]    
  \begin{center} 
      \includegraphics[width=0.47\textwidth]{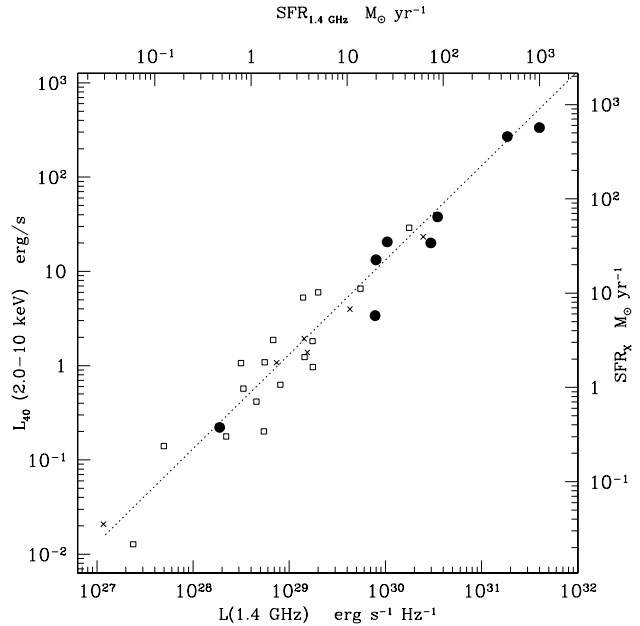}
  \end{center}
  \caption {\small The radio/X-ray luminosity relation holds in the Hubble 
   Deep Field too. Open symbols as in fig.\,(%
   \ref{pranalli-E3_fig:fig1}). Filled
   circles: deep sample.     Line:
                  best fit for local galaxies (eq.~%
   \ref{pranalli-E3:radiox}). The two galaxies with highest luminosity
   are both at $z\simeq 1.2$}
  \label{pranalli-E3_fig:fig3}
\end{figure}

On the other hand
the radio/FIR/hard X-ray relation is definitively linear.
In Fig.~(\ref{pranalli-E3_fig:fig2}) we plot 2--10 keV luminosities versus FIR
and radio ones. 
Least-squares fits yield:\\
\begin{eqnarray}
\Log (L_{\rm 2-10}) &= &(1.08\pm 0.09) ~\Log (L_{\rm FIR}) 
	-3.8\pm 0.3 \label{pranalli-E3:bf210a} \\
\Log (L_{\rm 2-10}) &= &(1.04\pm 0.11) ~\Log (L_{\rm 1.4})
	 +10.1\pm 3.4 \label{pranalli-E3:bf210b}
\end{eqnarray}\\
with mean $\delta$'s of 51\% and 69\%, respectively. 
The linearity and the dispersion are not significantly changed
neither by the inclusion of the supplementary sample 
($L_{2-10}\propto L_{\rm FIR}^{0.98\pm 0.07}$
and $L_{2-10}\propto L_{\rm 1.4}^{0.98\pm 0.07}$, $\delta=50\%$ and $57\%$ respectively),
nor by the use of the $60\mu$ luminosity ($L_{2-10}\propto L_{60\mu}^{1.00\pm 0.11}$).

It is also worth noticing that while the soft X-ray relations involve
some further uncertainties related to the possible presence of
intrinsic absorption, this is negligible in the 2--10 keV band for
column densities usually found in normal galaxies. 
A high column density ($N_{\rm H}\gtrsim 10^{22}$)
could partially obscure a possible AGN contribution;
however this seems unlikely for our sample because of the
selection of the objects by optical spectroscopy and of their moderate
X-ray luminosity ($<10^{42}$ erg/s).

The existence of a tight linear relation implies that the three
considered bands all carry the same information. Since the radio and
far infrared luminosity are indicators of the SFR, the 2--10 keV
luminosity should also be an indicator of SFR.

By repeating the best fit procedure under the assumption of strict
linear relationships, consistent within $1\sigma$ with the result of
eq.~(\ref{pranalli-E3:bf210a}) and (\ref{pranalli-E3:bf210b}),
we find:\smallskip
\begin{eqnarray}
\Log (L_{\rm 2-10}) &= &\Log (L_{\rm FIR}) - 3.51 \\
\Log (L_{\rm 2-10}) &= &\Log (L_{\rm 1.4}) + 11.12 
   \label{pranalli-E3:radiox}\qquad. 
\end{eqnarray}

From eq.~(\ref{pranalli-E3:eqsfr1}) and (\ref{pranalli-E3:eqsfr2}) we
propose:
\begin{equation}\label{pranalli-E3:eqsfrX}
{\rm SFR}=1.7\cdot10^{-40} ~L_{\rm 2-10 keV} \qquad\mbox{M$\sun$/yr}
\end{equation}
with $L_{\rm 2-10~keV}$ in erg/s.

\section{Star-forming galaxies in the Hubble Deep Field}

The 1~Ms {\em Chandra} (\cite{pranalli-E3:bran01}) and the 
radio 8.4 GHz (\cite{pranalli-E3:vla98}) catalogs
of the  Hubble Deep Field North (\object{HDF}) reach a
limiting flux which is sufficiently deep to detect star-forming
galaxies at redshifts up to $z\sim1.2$, and can be used to
check wether the radio/X-ray relation holds also for distant
galaxies.

Radio and X-ray positions were matched with an encircling radius of
$1.1^{\prime\prime}$.
All objects classified by \cite*{pranalli-E3:vla98} as AGN, or with an elliptical
or undefined morphology
were excluded; all other objects are accepted as ``candidate starbursts''.
Fluxes at 1.4\,GHz were calculated making use of 8.4\,GHz data and spectral indexes
measured for each object and reported in \cite*{pranalli-E3:vla98}.
Redshifts were taken from \cite*{pranalli-E3:cohen}. Fluxes in the 2--10 keV were
derived from a power-law fit to the observed 0.5--8 keV counts by
\cite*{pranalli-E3:bran01}. Radio and X-ray k-corrections were calculated for a
range of spectral indexes and redshifts; we chose to use mean k-corrections
since the induced error would not significantly affect the rest-frame fluxes.


We find that radio and X-ray luminosities of candidate
starburst galaxies  at larger redshifts follow the same relation
we observe in the
very local universe (Fig.~\ref{pranalli-E3_fig:fig3}), which
now spans 5 orders of magnitude.

\section{Conclusions}


We have analyzed a small but well defined sample of 17 star 
forming galaxies for which there is a homogeneous information
on optical, FIR, radio and X-ray bands (local sample). 
The BeppoSAX and ASCA X-ray data have been corrected for Galactic absorpion only. In 
agreement with a previous work (\cite{pranalli-E3:djf92}) we find that the 
logarithms of the soft (0.5--2 keV) X-ray luminosities are linearly 
correlated 
with the logarithms of either radio (1.4 GHz) and FIR luminosities, 
and that within the statistical errors these relationships are 
consistent with linearity between the corresponding luminosities. 
We have extended our analysis to the harder X-ray band, essentially 
free from internal absorption which may affect the soft X-ray fluxes, 
and found that there is a tight linear correlation between the X-ray 
luminosities in the 2--10 keV interval with both the radio and the FIR 
luminosities, normally assumed as the indicators of the star formation 
rate. We conclude that the origin of the hard X-ray emission must be
closely related to star formation.

 By selecting candidate starburst galaxies from the HDF North, whose 
redshift range extends up to $z\,\sim\,1.2$, we find that their X-ray ($L_{2-10}$)
and radio ($L_{1.4}$) luminosities closely follow the same linear
relationship derived for the local sample. This linear correlation
encompasses five orders of magnitude in luminosity, 
up to $L_{2-10} \sim$ several $10^{42}$ erg~s$^{-1}$ and a corresponding star 
formation rate $\sim 1000~ {\rm M}\sun$ yr$^{-1}$.
 We also notice that for the highest luminosity galaxies (which have $z\simeq 1.2$)
the observed 2--10 keV band corresponds to a rest-frame 4.5--22 keV, where the
emission is likely to be non-thermal. A plasma hot enough to dominate
the emission in this band would pose many problems 
in heating and bounding the gas.

 The way to understand the physics involved
in hard X-ray emission must go through a careful analysis of {\em Chandra}
and XMM-{\em Newton} observation, which have the necessary spatial resolution
and spectral sensitivity.
Moreover, since the radio/FIR relation
holds also locally in galaxies down to scales of about 100\,pc,
it is worth remarking that the two
star-bursting nuclei of \object{NGC\,3256} closely follow the
hard X-ray/radio relation, as suggested by our preliminary work on the
{\em Chandra} observation for this galaxy.
 The explanation of the radio/FIR correlation is still a matter of discussion;
we hope that these results may help in clarifying this issue.
 We are currently analysing possibilities for a theoretical interpretation
which, along with further constraints on this observational results, will
be the subject of a forthcoming paper.

\end{document}